\documentclass{elsart5}

 \usepackage{graphicx}

\usepackage{amssymb}

\begin{document}

\begin{frontmatter}
\title{Renormalization of Coulomb interactions in s-wave superconductor Na$_x$CoO$_2$}

\author[aff1]{Keiji Yada}
\author[aff1]{Hiroshi Kontani}
\address[aff1]{Department of Physics, Nagoya University, Furo-cho, Chikusa-ku, Nagoya, 464-8602, Japan}

\begin{abstract}
We study the renormalized Coulomb interactions due to retardation effect in Na$_x$CoO$_2$.
Although the Morel-Anderson's pseudo potential for $a_{1g}$ orbital $\mu^*_{a1g}$ is relatively large
because the direct Coulomb repulsion $U$ is large,
that for interband transition between $a_{1g}$ and $e_g'$ orbitals $\mu^*_{a1g,eg'}$ is very small
since the renormalization factor for pair hopping $J$ is square of that for $U$.
Therefore, the $s$-wave superconductivity due to valence-band Suhl-Kondo mechanism
will survive against strong Coulomb interactions.
The interband hopping of Cooper pairs due to shear phonons is essential
to understand the superconductivity in Na$_x$CoO$_2$.
\end{abstract}

\end{frontmatter}

Since the discovery of superconductivity in Na$_x$CoO$_2$, both theoretical and experimental studies have been actively performed to elucidate the mechanism of superconductivity and the pairing symmetry.
Various theoretical studies have suggested
that the development of magnetic fluctuation due to the existence of hole pockets composed of $e_g'$ orbitals
gives rise to an anisotropic superconductivity.
However, such hole pockets are not observed in ARPES measurements.
On the other hand, smallness of the impurity effect on $T_{\rm c}$ suggests that this system is
$s$-wave superconductivity due to electron-phonon interaction.
However, since a large repulsive force due to Coulomb interaction breaks $s$-wave Cooper pairs in transition-metal materials,
simple $s$-wave superconductivity may not be favored.

In our previous paper, we studied the multi-orbital $d$-$p$ model with electron-phonon interactions
and found that $A_{1g}$ mode phonons and $E_g$ mode phonons (we call breathing and shear phonons, respectively)
are strongly coupled with $t_{2g}$ electrons\cite{label2}.
Breathing phonons and shear phonons induce the intraband and the interband transition, respectively.
The interband transition of Cooper pairs due to shear phonons
enhances the effective attractive force as well as the transition temperature
even if $e_g'$-like bands are valence-band.
We call this mechanism the valence-band Suhl-Kondo (SK) mechanism.
It is highly possible that $s$-wave superconductivity is realized
because of the large attractive force against the Coulomb repulsion.
Shear phonons as well as breathing phonons are important to understand the electron-phonon mechanism
superconductivity in Na$_x$CoO$_2$.

In the present paper, we study the depairing effect due to Coulomb interactions
to confirm the reality of s-wave superconductivity in Na$_x$CoO$_2$.
We consider the pair hopping $J$ and the direct Coulomb repulsion $U$
because Na$_x$CoO$_2$ is a multi-orbital system.
The effective attractive force due to breathing phonons are reduced by $U$
and interband transition of Cooper pairs due to shear phonons are prevented by $J$.
In this paper, we calculate the renormalized Coulomb interactions due to retardation in Na$_x$CoO$_2$,
and we find out that the pair hopping $J$ is significantly reduced.

The energy scales of electron-phonon interactions and Coulomb interactions are of the order of $\omega_{\rm D}$ and $W$,
respectively, where $\omega_{\rm D}$ is the Debye frequency for phonons and $W$ is the band width.
In Na$_x$CoO$_2$, $\omega_{\rm D}$ for shear phonon and breathing phonon are approximately 500 and 600 cm$^{-1}$ ($\sim$0.06 and 0.07 eV),
respectively, and the $t_{2g}$ band width is $\sim1.5$ eV.
Coulomb interactions have larger energy scale, so they are reduced by retardation effect.
The renormalized Coulomb interactions are obtained by integrating high energy region.
\begin{eqnarray}
U^*_{a1g}&=&U-UAU^*_{a1g}-2JBJ^*_{a1g,eg'},\label{eq1}\\
U^*_{eg'}&=&U-UBU^*_{eg'}-JAJ^*_{a1g,eg'}-JBJ^*_{eg',eg'},\\
J^*_{a1g,eg'}&=&J-JBU^*_{eg'}-UAJ^*_{a1g,eg'}-JBJ^*_{eg',eg'},\\
J^*_{eg',eg'}&=&J-JBU^*_{eg'}-JAJ^*_{a1g,eg'}-UBJ^*_{eg',eg'},\label{eq4}
\end{eqnarray}
where $U^*_{a1g}$ ($U^*_{eg'}$) is the renormalized Coulomb repulsion for $a_{1g}$ ($e_g'$) orbital,
and $J^*_{a1g,eg'}$ ($J^*_{eg',eg'}$) is the renormalized pair hopping
between $a_{1g}$ orbital and $e_g'$ orbital ($e_g'$ orbitals).
$A$ and $B$ are the renormalization factors for $a_{1g}$ and $e_g'$ orbitals, respectively.
\begin{eqnarray}
A&=&\frac{T}{N}\sum_{\bf k}\sum_{\omega_{\rm D}<|\varepsilon_\ell|<W}|G_{a1g}({\bf k},{\rm i}\varepsilon_\ell)|^2,\label{eq5}\\
B&=&\frac{T}{N}\sum_{\bf k}\sum_{\omega_{\rm D}<|\varepsilon_\ell|<W}|G_{eg'}({\bf k},{\rm i}\varepsilon_\ell)|^2.\label{eq6}
\end{eqnarray}
We put $\omega_{\rm D}=550$ (cm$^{-1}$) for breathing and shear phonons, and $W=1.5$ (eV).
Na$_x$CoO$_2$ has $a_{1g}$-like band and two $e_g'$-like bands near Fermi level
and the tops of $e_g'$-like bands $\Delta$ are located below Fermi level.
So we assume that the density of states (DOS) in Na$_x$CoO$_2$ has step as shown in Fig. 1.
We put $\rho_{a1g}=0.38$, $\rho_{eg'}=0.65$,
which is derived from the value of DOS near Fermi level in the lattice model\cite{label1}.
  \begin{figure}[t]
\begin{center}
\includegraphics[scale =0.5]{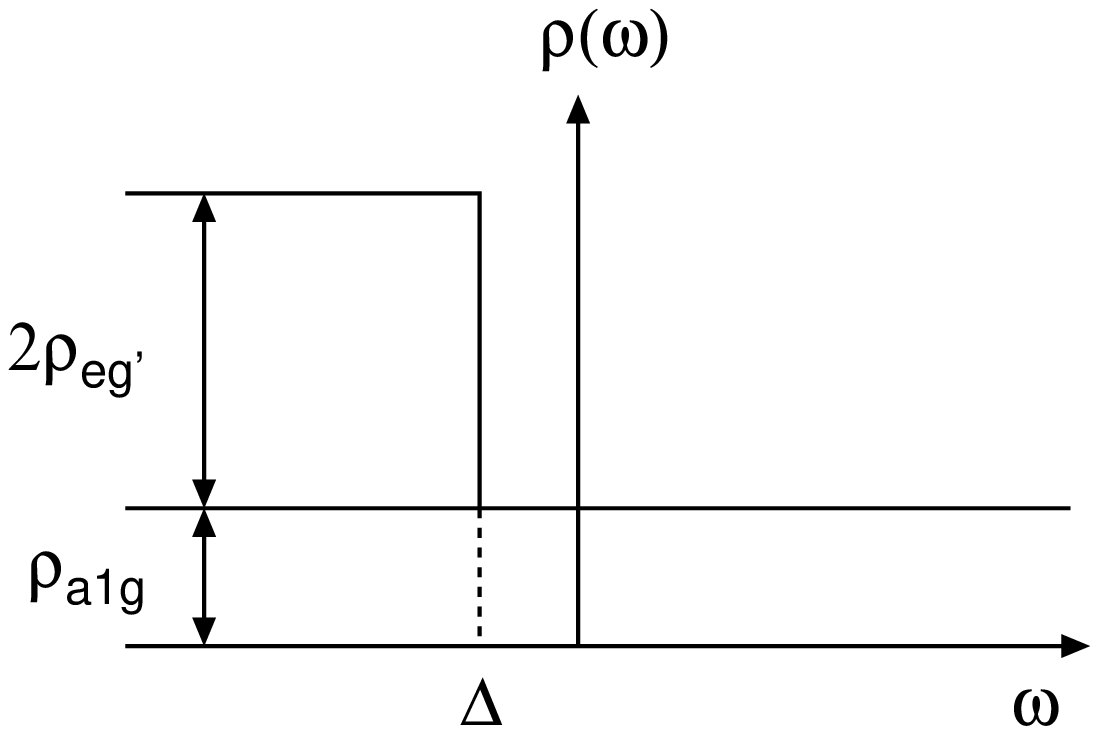}
\caption{DOS of Na$_x$CoO$_2$.}
  	\label{fig1}
\end{center}
  \end{figure}
Then, $A$ and $B$ are calculated from Eq. \ref{eq5} and Eq. \ref{eq6}, respectively,
$A=\rho_{a1g}\log(W/\omega_{\rm D})\approx 1.2, B=\rho_{eg'}\int_{\omega_{\rm D}}^W\frac{1}{\varepsilon}(\frac{1}{2}+\frac{1}{\pi}{\rm Arctan}(\frac{\Delta}{\varepsilon}))d\varepsilon$.
In the case of $\omega_{\rm D}\gg|\Delta|$, $B\approx\frac{1}{2}\rho_{eg'}\log(W/\omega_{\rm D})\approx1.0$.
In a realistic parameter regime $U\gg J$,
we can easily solve Eq. 1$\sim$4.
\begin{eqnarray}
U^*_{a1g}&\approx&U/(1+AU),\\
U^*_{eg'}&\approx&U/(1+BU),\\
J^*_{a1g,eg'}&\approx&J/(1+AU)(1+BU),\\
J^*_{eg',eg'}&\approx&J/(1+BU)^2.
\end{eqnarray}
Since the renormalization factor for $J$ is square of that for $U$,
$J$ is significantly renormalized.
When we put $U\approx W=1.5$ (eV), $U^*_{a1g}/U\approx1/2.8$ and $J^*_{a1g,eg'}/J\approx1/7$.
Fig. 2 shows the renormalized Coulomb interactions which is calculated from Eq. 1$\sim$4 for $U=10J=1.5$ (eV) and $U=4.5$, $J=0.3$ (eV).
The latter value of $U$ and $J$ is predicted by the ab-initio calculation\cite{bandcal}.
  \begin{figure}[t]
\begin{center}
\includegraphics[scale =0.42]{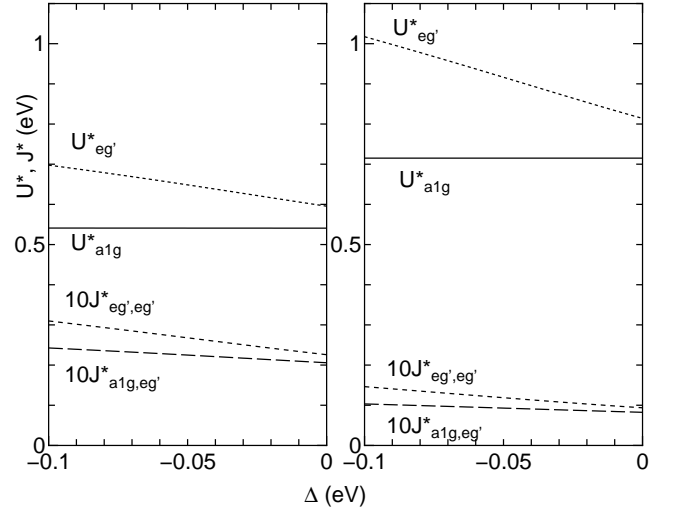}
\caption{Renormalized Coulomb interactions for $U=10J=1.5$ (eV) (left panel) and $U=4.5$, $J=0.3$ (eV) (right panel).}
  	\label{fig2}
\end{center}
  \end{figure}
The $\Delta$-dependence of $U^*_{a1g}$ is very small because the renormalization factor $A$ is constant.
However, $U^*_{eg'}$ and $J^*$ decrease as $\Delta$ increases.
This is because the renormalization factor $B$ increases with $\Delta$.
The Morel-Anderson's pseudo potential $\mu^*$ is obtained by multiplying $U^*$ by DOS\cite{morel};
$\mu^*=\mu/(1+\mu\log(W/\omega_{\rm D}))$, where $\mu=U\rho$.
In the case of $U=10J=1.5$ (eV),
$\mu^*$ for $a_{1g}$ $(e_g')$ orbital $\mu^*_{a1g}\approx0.2$ $(\mu^*_{eg'}\approx0.4)$.
$\mu^*$ for interband transition between $a_{1g}$ orbital and $e_g'$ orbital ($e_g'$ orbital) is
$\mu^*_{a1g,eg'}\approx0.01$ $(\mu^*_{eg',eg'}\approx0.02)$.
$\mu^*_{a1g}\approx0.2$ is relatively large value,
because $\mu^*$ is on the order of $0.1$ in normal metals.
On the other hand, $\mu^*_{a1g,eg'}\approx0.01$ is much smaller value.
So the effect of Coulomb interactions on interband hopping due to shear phonon is very small.

By using the same model shown in Fig. 1,
we derive the transition temperature for $\omega_{\rm D}\gg |\Delta|\gg T_{\rm c}$ ($\Delta<0$).
$T_{\rm c}\approx \omega_{\rm D}\exp(-1/\lambda^*_{\rm eff})$,
where $\lambda^*_{\rm eff}$ is given by
\begin{eqnarray}
\lambda^*_{\rm eff}=\lambda^*_1+\frac{2\lambda^*_2\lambda^*_3\{\frac{1}{2}\log(\frac{\omega_{\rm D}}{|\Delta|})+\frac{1}{\pi}\}}{1-\lambda^*_4\{\frac{1}{2}\log(\frac{\omega_{\rm D}}{|\Delta|})+\frac{1}{\pi}\}},\label{eq:lambda}
\end{eqnarray}
where $\lambda^*_1\sim\lambda^*_4$ are the attractive force induced by electron-phonon coupling.
The expression of $\lambda^*_1\sim\lambda^*_4$ are written in ref. \cite{label2}.
The second term in the right-hand side is the SK term.
Although relatively large $\mu^*_{a1g}$ and $\mu^*_{eg'}$ reduces $\lambda^*_1$ and $\lambda^*_4$,
$\lambda^*_2$ and $\lambda^*_3$ hardly change because $\mu^*_{a1g,eg'}$ takes a very small value.
Therefore, $\lambda^*_{\rm eff}$ can be positive value even in the case of $\lambda^*_1$ is negative value
because SK term remains a relatively large positive value.
This means that the superconductivity due to SK mechanism is possible
even if the attractive force due to breathing phonons cancel out by a large Coulomb repulsion.

In conclusion, we study the renormalized Coulomb interactions $U^*$ and $J^*$ in a multi orbital system
which describes the $t_{2g}$ band of Na$_x$CoO$_2$.
The pair hopping $J$ is significantly renormalized by the retardation effect,
and the effect of $J^*$ on interband hopping due to shear phonon is very small.
Therefore, valence-band SK mechanism works even if we consider the strong Coulomb interactions.
The interband hopping of Cooper pairs due to shear phonons is essential
in understanding the superconductivity in Na$_x$CoO$_2$.

\end{document}